\pdfoutput=1
\documentclass[runningheads]{llncs}
\usepackage{csquotes}
\usepackage[fleqn]{amsmath}
\usepackage{amssymb,amsfonts}
\usepackage{pifont}% http://ctan.org/pkg/pifont
\usepackage{algorithmic,algorithm}
\usepackage{mathtools}\usepackage{todonotes}
\usepackage{url}
\usepackage[utf8]{inputenc}
\usepackage{subfig}
\usepackage[para,online,flushleft]{threeparttable}
\usepackage{booktabs}
\usepackage{graphicx}
\usepackage{xcolor}
\usepackage{multirow}
\usepackage{booktabs}
\usepackage{bbding}
\usepackage[
    style=ieee,
    citestyle=numeric-comp,
    backend=biber,
    hyperref=true,
    maxnames=1,
    maxbibnames =3,
    minnames=1,
    mincrossrefs=1,
    natbib=true,
    sorting=none,
    sortcites=true,
    url=false,
    doi=false,
    eprint=false
]{biblatex}
\AtEveryBibitem{%
  \clearname{translator}%
  \clearlist{publisher}%
  \clearfield{pagetotal}%
\clearfield{pagetotal}%
\clearfield{issn}%
\clearfield{editor}%
\clearfield{month}%
\clearfield{notes}%
\clearfield{language}%

}
\DeclareFieldFormat
  [article,inbook,incollection,inproceedings,patent,thesis,unpublished]
  {title}{{#1}}

\usepackage[english]{babel}
\usepackage[colorlinks=true,urlcolor= blue]{hyperref}
\def\@fnsymbol#1{\ensuremath{\ifcase#1\or *\or \dagger\or \ddagger\or
   \mathsection\or \mathparagraph\or \|\or **\or \dagger\dagger
   \or \ddagger\ddagger \else\@ctrerr\fi}}
\makeatletter
\newcommand{\ssymbol}[1]{^{\@fnsymbol{#1}}}
\makeatother
\newcommand{\beginsupplement}{%
        \setcounter{table}{0}
        \renewcommand{\thetable}{S\arabic{table}}%
        \setcounter{figure}{0}
        \renewcommand{\thefigure}{S\arabic{figure}}%
      
     }

\newcommand{\Dataset}{\mathcal{D}}
\newcommand{\feature}{\mathbf{f}}

\newcommand{\Reals}{\mathbb{R}}

\newcommand{\zero}{\mathbf{0}}
\newcommand{\one}{\mathbf{1}}

\DeclareMathOperator{\EX}{\mathbb{E}}% expected value

\DeclareMathOperator*{\argmax}{arg\,max}

\newcommand{\ignore}[1]{} %for inline hidden comments

\definecolor{colornotes}{rgb}{0.6,0,0}

\newcommand{\image}{\boldsymbol{x}}

\newcommand{\Loss}{\mathcal{L}}

\newcommand{\bm}[1]{\mathbf{#1}}
\newcommand{\GaussianD}{\mathcal{N}}

\newcommand{\segmentation}{\mathbf{y}}

\newcommand{\prediction}{\mathbf{p}}

\newcommand{\datahref}[2]{\href{#1}{\textcolor{black}{#2}}}%

\begin{document}
\title{MaxStyle: Adversarial Style Composition for Robust Medical Image Segmentation}
\titlerunning{MaxStyle for Robust Medical Image Segmentation}
\author{Chen Chen\inst{1}(\Envelope), Zeju Li\inst{1}, Cheng Ouyang\inst{1}, Matt Sinclair\inst{1,2},  Wenjia Bai\inst{1,3,4}, Daniel Rueckert\inst{1,5}}
%insert a line % index{Last Name, First Name} for every author of the paper
% index{Chen, Chen}
% index{Li, Zeju}
% index{Ouyang, Cheng}
% index{Sinclair, Matt}
% index{Bai, Wenjia}
% index{Rueckert, Daniel}
\authorrunning{C. Chen et al.}
\institute{BioMedIA Group, Department of Computing, Imperial College London, UK
\and HeartFlow, USA
\and Data Science Institute, Imperial College London, UK
\and Department of Brain Sciences, Imperial College London, UK
\and Klinikum rechts der Isar, Technical University of Munich, Germany
\email{chen.chen15@imperial.ac.uk
}}

\maketitle            
                    \begin{abstract}
Convolutional neural networks (CNNs) have achieved remarkable segmentation accuracy on benchmark datasets where training and test sets are from the same domain, yet their performance can degrade significantly on unseen domains, which hinders the deployment of CNNs in many clinical scenarios. Most existing works improve model out-of-domain (OOD) robustness by collecting multi-domain datasets for training, which is expensive and may not always be feasible due to privacy and logistical issues. In this work, we focus on improving model robustness using a \emph{single-domain} dataset only. We propose a novel data augmentation framework called MaxStyle, which maximizes the effectiveness of style augmentation for model
OOD performance. It attaches an auxiliary style-augmented image decoder to a segmentation network for robust feature learning and data augmentation. Importantly, MaxStyle augments data with improved image style diversity and hardness, by expanding the style space with noise and searching for the worst-case style composition of latent features via adversarial training. With extensive experiments on multiple public cardiac and prostate MR datasets, we demonstrate that MaxStyle leads to significantly improved out-of-distribution robustness against unseen corruptions as well as common distribution shifts across \emph{multiple, different, unseen} sites and \emph{unknown} image sequences under both low- and high-training data settings. The code can be found at https://github.com/cherise215/MaxStyle.
\end{abstract}

\section{Introduction}
%% problem
Convolutional neural networks have demonstrated remarkable segmentation accuracy on images that come from the same domain (e.g. images from the same scanner at the same site), yet they often generalize poorly to unseen out-of-domain (OOD) datasets, which hinders the deployment of trained models in real-world medical applications. The performance degradation is mainly caused by the distribution mismatch between training and test data, often originating from the change of vendors, imaging protocols across different hospitals as well as imperfect acquisition processes which affect image quality. To alleviate this problem, a group of works focus on learning domain-invariant features with multi-domain datasets~\cite{Tao_2019_Radiology,liu2020ms,liu2020saml,Dou_2019_NIPS}. However, collecting and labeling such large datasets can be extraordinarily time-consuming, expensive, and not always feasible due to data privacy and ethical issues. In this work, we consider a more realistic yet more challenging setting: domain generalization from a single source only. 

A straightforward solution to improve domain robustness is data augmentation, which transforms and/or perturbs source domain data to resemble unseen data shifts~\cite{Wang_IJCAI_2021_DG_Survey,Xu_2021_ICLR_RandConv,Chen_2020_MICCAI_Adv_Bias,Chen_2021_latent_space_data_augmentation}. Existing works mainly consider directly perturbing the input space, which in general requires domain knowledge and expertise to design the perturbation function~\cite{Chen_2020_MICCAI_Adv_Bias}. Very few approaches investigate feature augmentation possibly due to the risk of perturbing semantic content of images. Perturbing features might change semantic attributes such as anatomical structures, which are essential for precise segmentation. An exception is MixStyle~\cite{Zhou_ICLR_2021_MixStyle}, which conducts content-preserving feature augmentation by randomly interpolating two samples' feature style statistics. However, there are several limitations with MixStyle: 1) difficulty in visualizing and interpreting augmented features in the high-dimensional space, 2) limited diversity of augmented styles due to the linear interpolation mechanism, 3) sub-optimal effectiveness for model robustness as it does not take the network's vulnerability into account at training. 

%% our contribution
In this work, we propose a novel data augmentation architecture, where an auxiliary image decoder is attached to a segmentation network to perform self-supervised image reconstruction and style augmentation. Such a design not only improves the interpretability of style augmented examples but also forces the network to learn auxiliary reconstructive features for improved out-of-domain robustness (Sec.~\ref{sec:auxiliary image decoder}). We further propose an enhanced style augmentation method: \textbf{MaxStyle}, which maximizes the effectiveness of style augmentation for model OOD performance (Sec.~\ref{sec:maxstyle}). Different from MixStyle, we expand the style space with additional noise and search for the worst-case style composition for the segmentor via adversarial training. Through extensive experiments, we show that MaxStyle achieves competitive robustness against corruptions and improves robustness against common distribution shifts across different sites and different image sequences (Sec.~\ref{sec:results}). 

\section{Related work}
\textbf{Style transfer}: Style transfer modifies the visual style of an image
while preserving its semantic content. Previous works on style transfer mainly focus on designing advanced generative adversarial networks (GAN) for improved image diversity and fidelity under the assumption that a large-scale dataset with diverse image styles is available~\cite{Karras_21_TPAMI_StyleGAN,Huang_2018_ECCV_Multimodal}. Recent findings show that domain shift is closely related to image style changes across different domains~\cite{Zhou_ICLR_2021_MixStyle,Li_2018_AdaBN} and can be alleviated by increasing the diversity of  training image styles ~\cite{Philip_CVPR_2019,Zhou_ICLR_2021_MixStyle,Yamashita_TMI_2021_Learning,Li_ICLR_2022_OOD,Wagner_MICCAI_2021_Structure,zhong2021advstyle}. One such successful example is MixStyle~\cite{Zhou_ICLR_2021_MixStyle}, which generates `novel' styles via simply linearly mixing style statistics from two arbitrary training instances from the same domain at feature level. DSU~\cite{Li_ICLR_2022_OOD} augments feature styles with random noise to account for potential style shifts. Compared with MixStyle and DSU, our proposed method, MaxStyle does not only span a \emph{larger} area in the feature style space but also covers \emph{harder} cases.\\
\textbf{Adversarial data augmentation:}
Adversarial data augmentation focuses on generating data samples that fool the network to predict the wrong class~\cite{Madry_2017_PGDattack,Goodfellow_2015_FGSM}. Such an approach can improve robustness against adversarial and natural corruptions~\cite{Madry_2017_PGDattack,Gilmer_ICML_2019}, and has shown promise for improved model generalization~\cite{Xie_CVPR_2020,Volpi_NeurIPS_2018_Generalizing,Qiao_CVPR_2020_MADA,Chen_2020_MICCAI_Adv_Bias,Chen_2021_latent_space_data_augmentation,zhong2021advstyle}. Most methods apply adversarial perturbation to the input space to improve unseen domain performance, e.g., adversarial noise~\cite{Miyato_2018_PAMI_VAT,Volpi_NeurIPS_2018_Generalizing,Qiao_CVPR_2020_MADA}, and adversarial bias field~\cite{Chen_2020_MICCAI_Adv_Bias}. Fewer explore feature space perturbation. Huang et al. proposed to adversarially drop dominant features (RSC) during training as a way to improve cross-domain performance~\cite{Huang_2020_ECCV_Self_challenging}. Chen et al. proposed to project masked latent features back to input space in an adversarial manner for improved interpretability and effectiveness~\cite{Chen_2021_latent_space_data_augmentation}. Our method follows this feature-to-input space data augmentation mechanism, but restricts the adversarial perturbation at the style level, generating \emph{style-varying, shape-preserving} hard examples.

%% GAN-based data augmentation
% \textbf{Style transfer}. 
% Previous work has attempted to alleviate the problem via directly perturb image with noises or partly maskin features for robust feature learning. In this work, we consider to apply feature style augmentation, i.e. MixStyle, with the appealing benefit of preserving shape features while increasing the variations of styles. 

\section{Method}
% %% Problem set up
% \textbf{Problem definition:} Let $\imagespace$ be the input image space and $\labelspace$ the label space and $\classifier_{seg}(\cdot;\theta):\imagespace \rightarrow \labelspace$ a parameterised segmentation model to learn a mapping from $\imagespace$ to $\labelspace$. Given a set of labelled images $\{(\image, \segmentation)^{(i)}\}_{i=1...n}$ collected from a single source domain $\Domain^{S}_{train}$  for training, our goal is to learn an optimal segmentation model $\classifier_{seg}({\cdot;\theta^*})$ so that it can perform well not only on the intra-domain test data $\Domain^{S}_{test}$ but also on out-of-domain test datasets from test domains $T$: $\{\Domain_{test}^{T}\}$ with unknown domain shifts without dramatic failures. 

\subsection{Preliminaries: MixStyle}
\begin{figure}[t]
    \centering   
    \includegraphics[width=0.9\textwidth]{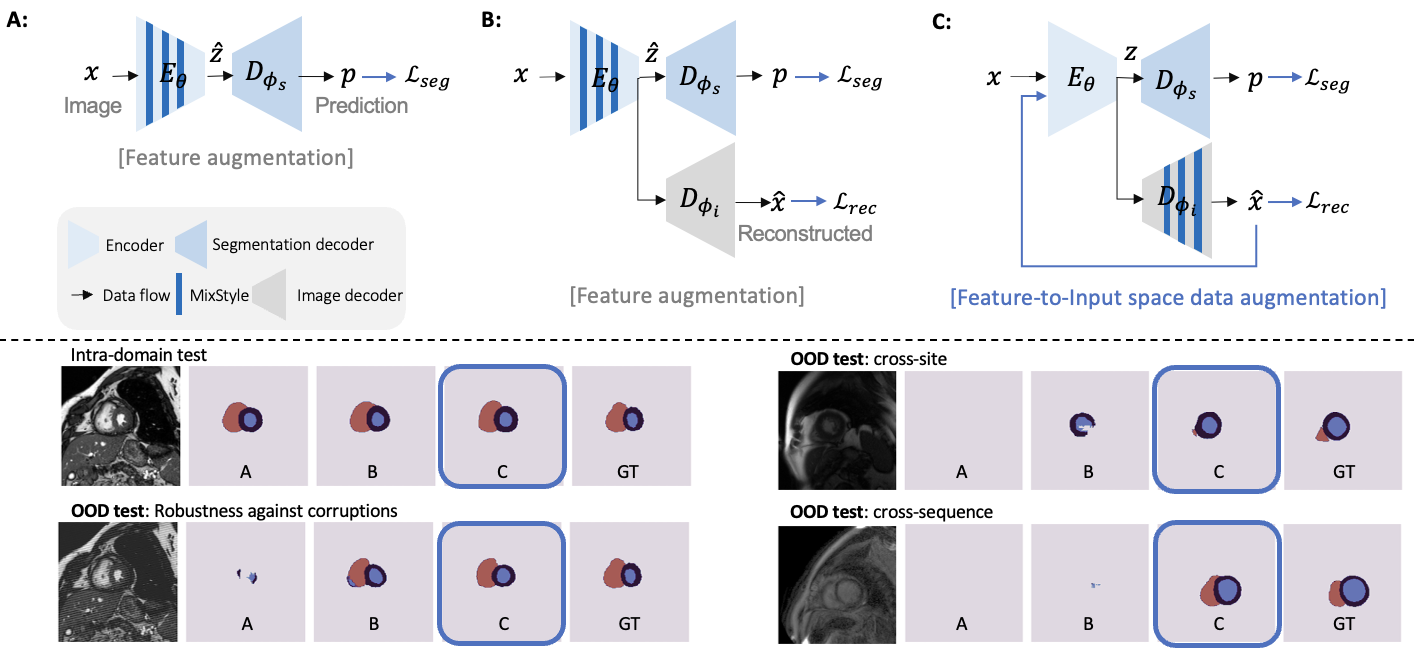}
    \caption{A) the original use of MixStyle as feature augmentation-based  regularization method with a standard encoder-decoder structure. B) MixStyle applied to regularize a dual-branch network with an auxiliary image decoder $D_{\phi_i}$ attached for image reconstruction. C) We propose to apply MixStyle in the auxiliary  decoder $D_{\phi_i}$ to generate stylized images for feature-to-input space data augmentation instead (MixStyle-DA), which leads to improved model robustness across different OOD test data, compared to A, B (sebottom). GT: ground-truth.}
    \label{fig:mix_style_variants}
\end{figure}

As mentioned before, MixStyle~\cite{Zhou_ICLR_2021_MixStyle} is a feature augmentation method for improving model generalization and robustness, which perturbs feature style statistics to regularize network training. Formally, let $\feature_i, \feature_j  \in \Reals^{c \times h \times w} $ be $c$-dimensional feature maps  extracted at a certain CNN layer for image $\image_i$ and image $\image_j$, respectively. 
MixStyle performs style augmentation for $\feature_i$ by first normalizing it with its channel-wise means and standard deviations $\mu(\feature_i), \sigma(\feature_i) \in \Reals^{c}$ : $\overline{\feature}_{i} = \frac{\feature_i-\mu(\feature_i)}{\sigma(\feature_i)}$ and then transforming the feature with a linear combination of style statistics $i: \{\sigma(\feature_i),\mu(\feature_i)\}$ and $j: \{\sigma(\feature_j),\mu(\feature_j)\}$. This process can be defined as:
\begin{small}
\begin{multline}
\operatorname{MixStyle}(\feature_i)=\boldsymbol{\gamma}_{mix} \odot \overline{\feature}_{i}+{\boldsymbol{\beta}_{mix}},\\
\boldsymbol{\gamma}_{mix} = \lambda_{mix}\sigma(\feature_i) +(1-\lambda_{mix})\sigma(\feature_j),\,
\boldsymbol{\beta}_{mix} = \lambda_{mix}\mu(\feature_i) +(1-\lambda_{mix})\mu(\feature_j),
\label{eq:mix_style}
\end{multline}
\end{small}where $\odot$ denotes element-wise multiplication; $\lambda_{mix}$ is a coefficient controlling the level of interpolation, randomly sampled from $[0,1]$. Acting as a plug-and-play module, MixStyle can be inserted between any CNN layers. It is originally designed as an explicit regularization method for a standard encoder-decoder structure, which perturbs shallow features (e.g., features from the first three convolutional blocks) in an image encoder $E_\theta$ parametersied by $\theta$, see Fig.~\ref{fig:mix_style_variants}. A segmentation decoder $D_{\phi_s}$ then performs prediction with the perturbed high-level representation $\hat{\boldsymbol z}$. The whole network is optimized to minimize the segmentation loss $\Loss_{seg}(D_{\phi_s}(\hat{\boldsymbol z}),\segmentation)$, supervised by the ground-truth $\segmentation$.
\subsection{Robust feature learning and improved interpretability with auxiliary image decoder}
\label{sec:auxiliary image decoder}
In this work, for improved OOD robustness and better interpretability of style augmentation, we propose to adapt the standard encoder-decoder to a dual-branch network with an auxiliary image decoder $D_{\phi_i}$ attached, see Fig.~\ref{fig:mix_style_variants}B,C. The network is supervised with an additional image reconstruction loss $\Loss_{rec}(\hat\image,\image)$, allowing itself to exploit both complementary image content features and task-specific shape features for the segmentation task~\cite{Chen_2021_latent_space_data_augmentation}. 
% Reconstructive features has been found to be critical for OOD detection and robust inference in traditional machine learning algorithms~\cite{Fidler_2006_TPAMI_Combining}. 
We further propose to insert the style augmentation layers in the image decoder $D_{\phi_i}$ (Fig.~\ref{fig:mix_style_variants}C) rather than in the encoder $E_\theta$ (Fig.~\ref{fig:mix_style_variants}B), allowing to generate diverse stylized images with the same high-level semantic features $\boldsymbol z$ for \emph{direct} data augmentation. Such a design also improves the interpretability of feature style augmentation. More importantly, our experimental results show that $C>B>A$ in terms of OOD robustness, see Fig.~\ref{fig:mix_style_variants} (bottom) and Fig. S1 (supplementary). The segmentation network trained w/ approach C is more robust against corruptions and unseen domain shifts across different sites and different image sequences. We name this new method as `MixStyle-DA' to differentiate it from the original one. We believe this data augmentation-based approach is preferred for model robustness as it tackles model over-fitting from the root, limited training data. 

% \noindent\textbf{Connection to cognitive science.} The network can be viewed as a learner, taught to perform tasks in different ways. In this case, B is a passive approach as the learner is taught to ignore style relevant perturbations even before it fully understands the high-level concept. By contrast, C requires the learner to first understand the basic concept, and then infer as well as connect to other concepts to challenge themselves. Such an active learning approach is preferred as it promotes higher-order thinking with improved learning efficacy~\cite{minhas2012effects}. 

\subsection{MaxStyle}
\label{sec:maxstyle}
\begin{figure}[t]
    \centering   
    \includegraphics[width=0.89\textwidth]{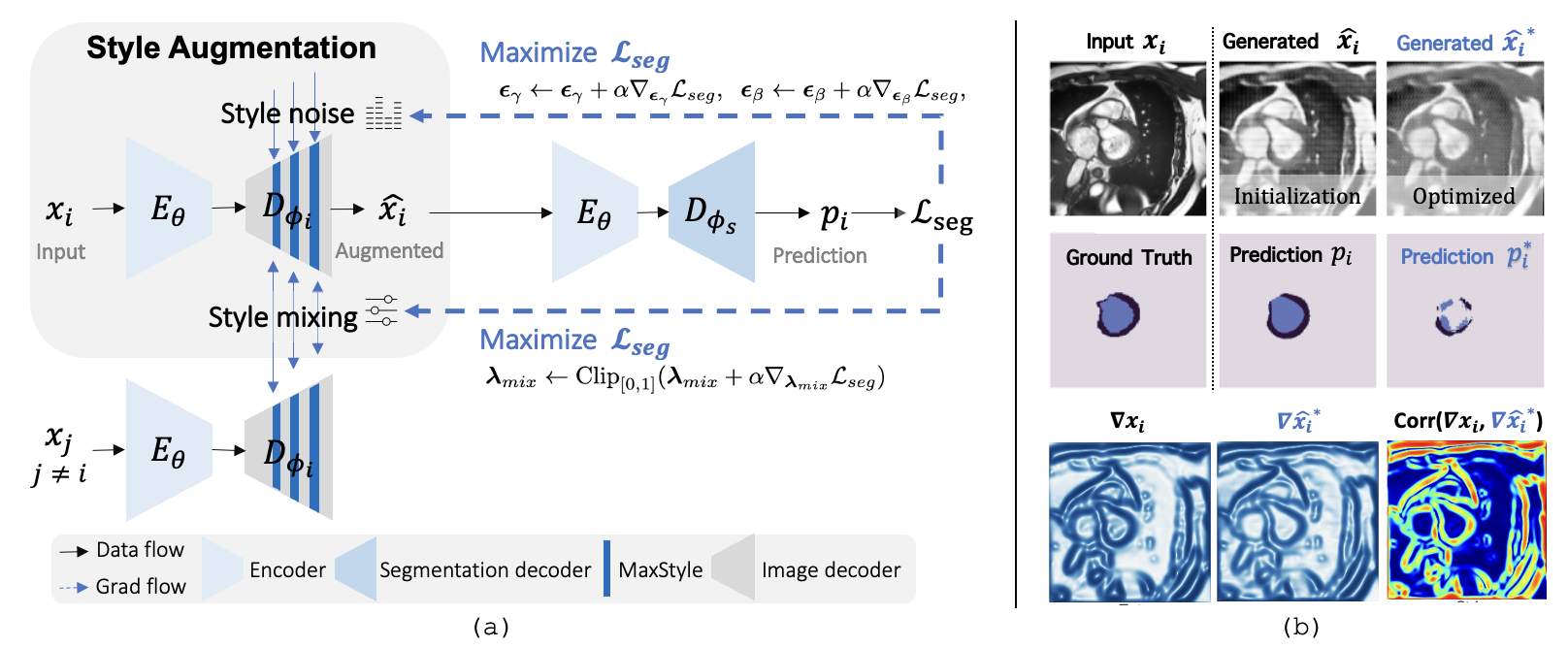}
    \caption{\textbf{MaxStyle} overview. a) MaxStyle reconstructs $\image_i$ with augmented feature styles via style mixing and noise perturbation in the image decoder $D_{\phi_i}$. Adversarial training is applied, in order to search for `harder' style composition to fool the segmentation network ($E_\theta \circ D_{\phi_s}$). b) MaxStyle  generates a style-optimized image $\hat\image^*$, which fools the network to under-segment ($\hat\prediction^*$). The anatomical structures remain almost unchanged with \emph{high} correlation (Corr) between two images' gradient fields:  $\nabla{\image}$, $\nabla\hat{\image}^*$.}
    \label{fig:maxstyle}
\end{figure}
On top of the dual-branch architecture with the data augmentation strategy presented above (Fig.~\ref{fig:mix_style_variants}C), we propose MaxStyle, a novel method that consists of both style mixing and noise perturbation, improving the diversity and sample hardness. As visualized in Fig.~\ref{fig:maxstyle}(a), we introduce additional style noise to expand the style space and apply adversarial training in order to find optimal linear coefficients as well as to generate adversarial style noise for effective data augmentation. Specifically, given feature $\feature_i$ extracted at a certain CNN layer in the image decoder $D_{\phi_i}$ with $\image_i$ as input, MaxStyle augments $\feature_i$ via:
\begin{align}
&\operatorname{MaxStyle}(\feature_i)=(\boldsymbol{\gamma}_{mix}+\Sigma_{\gamma}\cdot \boldsymbol\epsilon_{\gamma}) \odot \overline{\feature}_{i} + (\boldsymbol\beta_{mix}+\Sigma_{\beta}\cdot \boldsymbol\epsilon_{\beta}),\;
\end{align}
where the normalized feature $\overline{\feature_i}$ is transformed with mixed styles $\boldsymbol\gamma_{mix},\boldsymbol\beta_{mix}$ (Eq.~\ref{eq:mix_style}) plus additional style noise $\Sigma_{\gamma}\cdot \boldsymbol\epsilon_{\gamma},\Sigma_{\beta}\cdot \boldsymbol\epsilon_{\beta} \in \Reals^c$ to better explore the space of possible unknown domain shifts. In order to bound the style noise within a reasonable range, similar to \cite{Li_ICLR_2022_OOD}, we sample the style noises from a re-scaled Gaussian distribution with variance  $\Sigma_{\gamma}, \Sigma_{\beta} \in \Reals^c$ estimated from a batch of $B$ instances' style statistics (including $\image_i$): $\Sigma_{\gamma} = \sigma^2(\{\sigma(\feature_j)\}_{j={1...i,...B}})\;, \Sigma_{\beta} = \sigma^2(\{\mu(\feature_j)\}_{j={1...i,...B}}), \boldsymbol\epsilon_\gamma,\boldsymbol\epsilon_\beta \sim \GaussianD(\zero,\one)$
~\footnote{The re-parameterization trick is applied here for ease of follow-up optimization.}. \\
\noindent\textbf{Style optimization via adversarial training:}
The generated noise $\boldsymbol\epsilon_{\gamma},\boldsymbol\epsilon_{\beta}$ and the style mixing coefficient $\lambda_{mix}$ are then updated by maximizing the segmentation loss $\Loss_{seg}$ so that the synthesized image $\hat{\image} = D_{\phi_i}(E_\theta(\image);\boldsymbol\epsilon_\gamma,\boldsymbol\epsilon_{\beta},\lambda_{mix})$\footnote{For simplicity, we omit non-learnable parameters such as the sampling operator to choose instance $\image_j$ from a batch for style mixing.} could fool the network to produce an incorrect prediction $\hat\prediction=D_{\phi_s}(E_\theta(\hat{\image}))$. Gradient ascent is employed to update the underlying style-related parameters:
\begin{small}
\begin{align}
&\boldsymbol\epsilon_\gamma \leftarrow \boldsymbol\epsilon_\gamma+\alpha \nabla_{\boldsymbol\epsilon_\gamma}{\Loss_{seg}(\hat\prediction,\segmentation)}, \;\;
\boldsymbol \epsilon_\beta \leftarrow \boldsymbol\epsilon_\beta+\alpha \nabla_{\boldsymbol\epsilon_\beta}\Loss_{seg}(\hat\prediction,\segmentation),\\
&\boldsymbol\lambda_{mix} \leftarrow \operatorname{Clip}_{[0,1]}( \boldsymbol\lambda_{mix}+\alpha \nabla_{\boldsymbol\lambda_{mix}}{\Loss_{seg}(\hat\prediction,\segmentation)}).
\end{align}
\end{small}
Here $\alpha$ denotes the step size. We clip the value of $\lambda_{mix}$ to ensure it lies in $[0,1]$.  To summarize, training with MaxStyle augmentation can be then written as minimax optimization using the training set $\Dataset$: 
\begin{equation}
\begin{small}
\begin{aligned}
\min_{\theta,\phi_i,\phi_s} & \EX_{\image,\segmentation \sim \Dataset}  \underbrace{\Loss_{seg}(D_{\phi_s}(E_\theta({\image})),\segmentation)
     +\Loss_{rec}(D_{\phi_i}(E_\theta(\image)),\image)}_{\textcolor{gray}{\image\; \small{\textit{as input for network optimization}}}}\\
&+\underbrace{\Loss_{seg}(D_{\phi_s}(E_\theta(\hat{\image}^*)),\segmentation) +\Loss_{rec}(D_{\phi_i}(E_\theta(\hat{\image}^*)),\image)}_{\textcolor{gray}{\hat\image^* \; \small{\textit{as input  for network optimization}}}}\\
s.t.\; &\lambda_{mix}^{*},\boldsymbol\epsilon_\gamma^{*},{\boldsymbol\epsilon_\beta}^{*}  = \argmax  \Loss_{seg}(\hat\prediction,\segmentation){\textcolor{gray}{\scriptsize{\textit{\;\;(adversarial style optimization)}}}}.
\end{aligned}
\end{small}
\end{equation}
\setlength\intextsep{0pt}The whole network is optimized using both input image $\image$ and its style augmented image $\hat\image^*$, minimizing the multi-task loss: segmentation loss $\Loss_{seg}$ and image reconstruction loss $\Loss_{rec}$. Here $\hat{\image}^*= D_{\phi_i}(E_{\theta}(\image);\lambda_{mix}^{*},\boldsymbol\epsilon_\gamma^{*},\boldsymbol\epsilon_\beta^{*})$ is generated using optimized style parameters. An example of MaxStyle generated images is shown in Fig.~\ref{fig:maxstyle}b. The network is fooled to under-segment the style-augmented image $\hat\image^*$ although the underlying target structures remain almost the same. More examples are provided in the supplementary. Of note, MaxStyle is suitable for general segmentation networks thanks to its plug-and-play nature and the attached decoder can be removed at test time.

\section{Experiments and results}
\subsection{Data: Cardiac MR segmentation datasets}
a) Single source domain: The public \datahref{https://www.creatis.insa-lyon.fr/Challenge/acdc/}{ACDC} dataset~\cite{Bernard_2018_TMI} is used for network training and intra-domain test, which contains 100 subjects (bSSFP sequences) collected from a \textbf{single} site. We randomly split the dataset into 70/10/20 for training/validation/intra-domain test.To simulate common clinical scenarios where the training data is far less than testing data, we randomly selected 10 subjects (10 out of 70) for training and repeated experiments three times. We also validated our method using all 70 training subjects  following~\cite{Chen_2021_latent_space_data_augmentation}. \\
b) Multiple unseen test domains:  We collated several
\emph{public} datasets where each capture different types of 
distribution shift or quality degradation for comprehensive OOD performance evaluation. They are 1) \textit{ACDC-C} for robustness against imaging artefacts evaluation. Following~\cite{Chen_2021_latent_space_data_augmentation}, we synthesized a corrupted version of the ACDC test set where each test image has been randomly corrupted by one of \textbf{four, different} types of MR artefacts, three times for each artefact type,  using \datahref{https://github.com/fepegar/torchio}{TorchIO} toolkit~\cite{perez-garcia_torchio_2021}; \textit{ACDC-C} contains $240\;(20\times3\times 4)$ subjects in total; 2) a large \emph{cross-site} test set with real-world distribution variations. It consists of 195 subjects (bSSFP sequences) across \textbf{{four, different}} sites, which are from two public challenge datasets: \datahref{https://www.ub.edu/mnms/}{M\&Ms}~\cite{Campello_2021_TMI_Multi}, \datahref{http://www.sdspeople.fudan.edu.cn/zhuangxiahai/0/mscmrseg19/}{MSCMRSeg}~\cite{Zhuang_Arxiv_2020}; 3) \textit{cross-sequence} test set, which consists of 45 LGE sequences from \datahref{http://www.sdspeople.fudan.edu.cn/zhuangxiahai/0/mscmrseg19/}{MSCMRSeg}~\cite{Zhuang_Arxiv_2020}. Detailed information can be found in the Supplementary. We employed the image pre-processing and standard data augmentation (incl. common photo-metric and geometric image transformations) described in~\cite{Chen_2021_latent_space_data_augmentation} as our default setting.

% \textbf{Prostate segmentaion datasets.}
% \emph{a) single source domain:} The public prostate segmentation dataset from the Medical Decathlon challenge~\footnote{\url{http://medicaldecathlon.com/}} is used for training and intra-domain test, which contains 32 labelled prostate multiparametric MRI cases (T2) collected from the Radboud University Medical Center (RUNMC). Specifically, we randomly split the dataset into 22/3/7 for training/validation/intra-domain test; \emph{b) Multi-site unseen test domains:}  A large public prostate multi-site dataset~\footnote{\url{https://liuquande.github.io/SAML/}} is used for cross-domain generalization test, where images were collected from \emph{six} different data sources out of three public datasets~\cite{liu2020ms,liu2020saml}. Detailed information can be found in Table~\ref{tab:prostate multi-site robustness test sets}.
\subsection{Implementation and experiment set-up}
We adopted the dual-branch network presented in~\cite{Chen_2021_latent_space_data_augmentation} as our backbone, which has demonstrated state-of-the-art robustness on the cardiac segmentation task. We applied MaxStyle layers to shallow features  in $D_{\phi_i}$ for optimal performance, i.e. inserted after each of the last three convolutional blocks. Each layer is activated at a probability of 0.5 and batch features at each layer are randomly shuffled for style mixing to trade-off diversity and strength as suggested by~\cite{Zhou_ICLR_2021_MixStyle}. For the adversarial style optimization, we randomly sample $\lambda_{mix}$ from the uniform distribution $\mathcal{U}_{[0,1]}$ and $\boldsymbol\epsilon_{\gamma,\beta}$ from $\GaussianD(\zero,\one)$. Adam optimizer was adopted for both style optimization and network optimization. We empirically set $\alpha=0.1$ with 5 iterations, yielding sufficient improvement. For network optimization, we set learning rate $=1e^{-4}$ and batch size $=20$, following~\cite{Chen_2021_latent_space_data_augmentation}. For cardiac small training set, we trained the network for 1,500 epochs to ensure convergence. For larger training data, 600 epochs were sufficient. We employed mean squared error loss for $\Loss_{rec}$ and  cross-entropy loss for $\Loss_{seg}$. 
% Experiments were performed on an Nvidia$^{\tiny{\text{\textregistered}}}$ GPU, using Pytorch. The average Dice score is reported for segmentation performance evaluation.
We compare our method with the \emph{baseline} (w/o MaxStyle augmentation) and competitive random and adversarial augmentation methods using the same network and the same multi-task loss, including: a) input space augmentation:  RandConv~\cite{Xu_2021_ICLR_RandConv}, adversarial noise (Adv Noise)~\cite{Miyato_2018_PAMI_VAT}, and adversarial bias field (Adv Bias)~\cite{Chen_2020_MICCAI_Adv_Bias}; b) feature augmentation: RSC~\cite{Huang_2020_ECCV_Self_challenging}, MixStyle~\cite{Zhou_ICLR_2021_MixStyle}, DSU~\cite{Li_ICLR_2022_OOD}; c) feature-to-input space data augmentation via decoder $D_{\phi_i}$: latent space masking (LSM)~\cite{Chen_2021_latent_space_data_augmentation}. All  methods were implemented in PyTorch with their recommended set-ups. 

\subsection{Results}
\label{sec:results}
\begin{figure}[t]
    \centering   
    \includegraphics[width=0.8\textwidth]{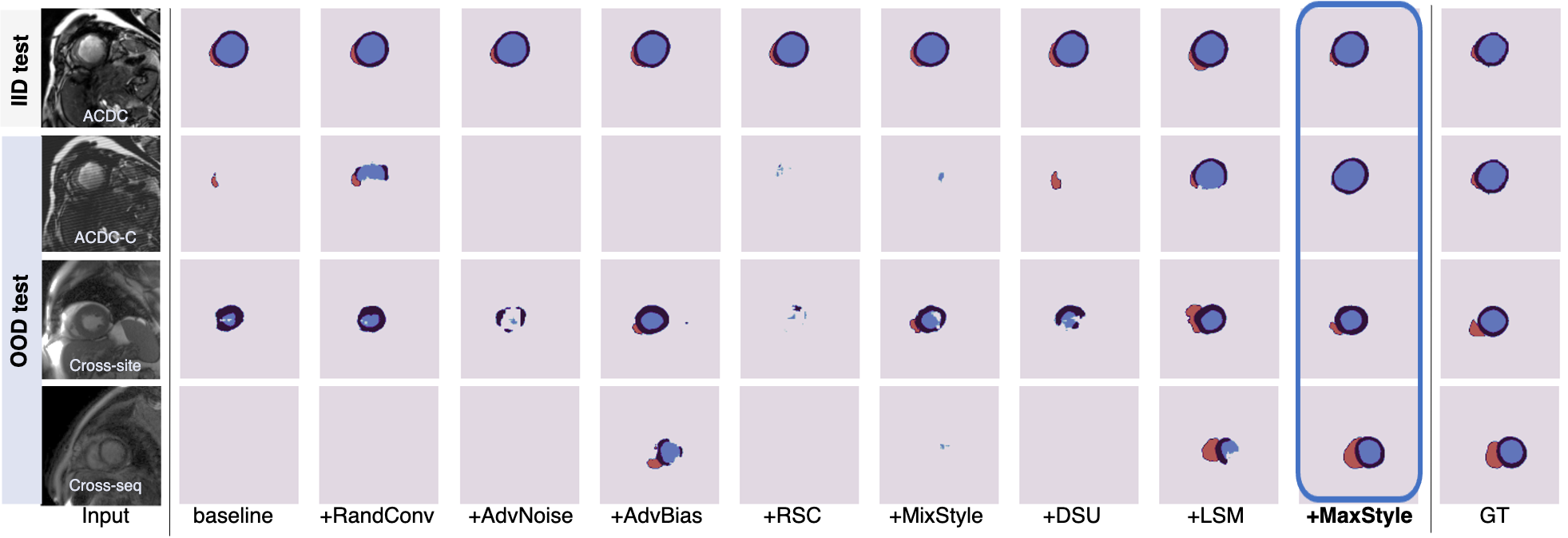}
    \caption{Qualitative results under the cardiac low-data regime. GT: ground-truth.}
    \label{fig:result_visualization}
\end{figure}

Results of models trained with cardiac low data regime and high data regime are provided Table~\ref{tab:cardiac_low_high}. We also plot the segmentation results in Fig.~\ref{fig:result_visualization} for visual comparison.
From the
results, we see that MaxStyle outperforms \emph{all} baseline methods across \emph{all} OOD test sets with different types of domain shifts, providing the largest improvements in both low- and high-data training regimes (+25\%, +11\%). In particular, MaxStyle significantly improves the segmentation performance on the most challenging cross-sequence dataset with the largest domain shift.  By generating images with challenging styles during training, MaxStyle forces the model to be more shape-biased rather than texture/style-biased, which is beneficial for model robustness~\cite{Geirhos_2019_ICLR_BiasedTexure}. To provide additional insights of MaxStyle, we also provide t-SNE visualization of latent feature embeddings in the supplementary. We also found that the model trained w/ {MaxStyle} using 10 ACDC subjects even achieves higher OOD performance than the \emph{baseline} model trained w/ 70 subjects (0.7420 vs 0.7287), suggesting MaxStyle's superior data efficiency. In addition to cardiac segmentation, we also validate our method on public prostate segmentation datasets~\cite{liu2020ms,Antonelli_2021_Medical_Decathlon,liu2020saml,NCI-ISBI_2013,Lemaitre_2015_I2CVB,Litjens_2014_Promise_12} to test the generality of our method for medical image segmentation. MaxStyle also achieves the top OOD performance against the competitive methods across \emph{six, different, unseen} test sites. Datasets details and results are shown in the supplementary.

\begin{table}[t]
\centering
\caption{Cardiac segmentation performance across multiple unseen test sets. IID: \emph{ACDC} test set performance. OOD: average performance across unseen OOD test sets: \textit{ACDC-C}, \textit{cross-site}, and \textit{cross-sequence}. Reported values are average Dice scores across different test sets. }
\label{tab:cardiac_low_high}
\resizebox{0.8\textwidth}{!}{%
\begin{threeparttable}
\begin{tabular}{@{}l|c|c|rrr|c|c|rrr@{}}
\toprule
\multirow{2}{*}{Method} & \multicolumn{5}{c|}{Low-data regime (10 training subjects)} & \multicolumn{5}{c}{High-data regime (70 training subjects)} \\ \cmidrule(l){2-11} 
 & IID & OOD & \textit{ACDC-C} & \textit{Cross-site } & \textit{Cross-sequence} & IID & OOD & \textit{\textit{ACDC-C}} & \textit{\textit{Cross-site }} & \textit{\textit{Cross-sequence}} \\ \midrule
\emph{baseline} & 0.8108 & 0.5925~\textcolor{gray}{(-)} & 0.6788 & 0.6741 & 0.4244 & 0.8820 & 0.7287~\textcolor{gray}{(-)} & 0.7788 & 0.8099 & 0.5974 \\ \midrule
+RandConv & 0.8027 & 0.6033~\textcolor{gray}{(+2\%)} & 0.6872 & 0.6717 & 0.4510  &0.8794 & 0.7371~\textcolor{gray}{(+1\%)} & 0.7967 & 0.7747 & 0.6399 \\
+Adv Noise & 0.8080 & 0.6104~\textcolor{gray}{($+3\%$)} & 0.6868 & 0.6804 & 0.4641 & 0.8852 & 0.7291~\textcolor{gray}{(+0\%)} & 0.7808 & 0.8095 & 0.5970 \\
+Adv Bias & 0.8114 & 0.6594~\textcolor{gray}{(+11\%)} & 0.6937 & 0.7517 & 0.5327 & {0.8865} & 0.7450~\textcolor{gray}{(+2\%)} & 0.7839 & 0.8250 & 0.6262 \\\midrule
+RSC & {0.8169} & 0.6252~\textcolor{gray}{(+6\%)} & 0.7010 & 0.7015 & 0.4729 & 0.8844 & 0.7379~\textcolor{gray}{(+1\%)}& 0.7757 & 0.8115 & 0.6267 \\
+MixStyle & 0.8024 & 0.6263~\textcolor{gray}{(+6\%)} & 0.7002 & 0.6984 & 0.4804 & 0.8788 & 0.7651~\textcolor{gray}{(+5\%)}& 0.7961 & 0.8130 & 0.6863  \\
+DSU & 0.8127 & 0.6088~\textcolor{gray}{(+3\%)} & 0.6972 & 0.7008 & 0.4285 & 0.8784 & 0.7553~\textcolor{gray}{(+4\%)} & 0.8031 & 0.8126 & 0.6503 \\\midrule
+LSM & 0.7899 & 0.6462~\textcolor{gray}{(+9\%)} & 0.7221 & 0.7015 & 0.5149 & 0.8754 & 0.7579~\textcolor{gray}{(+4\%)} & 0.8213 & 0.8025 & 0.6500\\
% \textbf{MixStyle-DA} & \textbf{0.8222} & 0.7231~\textcolor{gray}{(+22\%)} & 0.7483 & 0.7758 & 0.6453 & & 0.~\textcolor{gray}{(+\%)} & 0. & 0.\\
% \textbf{MaxStyle (soft)} & \textbf{0.8178} & \textbf{0.7461}~\textcolor{gray}{(+26\%)} & \textbf{ 0.7789} & \textbf{0.7757} & \textbf{0.6836}  & 0.8796 & \textbf{0.8111}~\textcolor{gray}{(+11\%)} & \textbf{0.8435} & \textbf{0.8266} & \textbf{0.7632} \\ 
+\textbf{MaxStyle} & $0.8104^\sim$ & \textbf{0.7420~\textcolor{gray}{(+25\%)}}$\ssymbol{5}$& \textbf{0.7745} & \textbf{0.7645}& \textbf{0.6869}&
0.8727$\ssymbol{4}$ & \textbf{0.8125~\textcolor{gray}{(+11\%)}}$\ssymbol{5}$ & \textbf{0.8408} & \textbf{0.8294} & \textbf{0.7673}\\
\bottomrule
\end{tabular}%
\begin{tablenotes}
\item[$\sim$]:p-value$>0.5$ \item[$\ssymbol{4}$] p-value$>1e^{-4}$ 
\item[$\ssymbol{5}$] p-value$\ll 1e^{-4}$ (compared to Baseline results)
\end{tablenotes}
\end{threeparttable}
}
\end{table}
\begin{table}[t]
\centering
\caption{The effect of expanding style search space (style noise \& style mixing) and applying adversarial style optimization (AdvOpt) for domain generalization. Reported values are mean (std) of Dice scores across all \emph{ten} cardiac test (sub)sets. Experiments were performed under the cardiac high-data training setting. }
\label{tab:abl_adversarial_training}
\resizebox{0.9\textwidth}{!}{%
\begin{tabular}{@{}c|c|c|c|c|c@{}}
\toprule
%  rand style mixing \& adv training & w/o style mixing \& adv training &  w/o adv training  &  w/ adv training \\ \midrule
%  0.7293 (0.0810) & 0.7607 (0.0515) & - (-) & 0.7721 (0.0428)
% w/o Style Noise & w/o Style Mixing & w/o Adv Style Optimization & w/o Adv Style Optimization \& Style noise (MixStyle-DA) & \textbf{MaxStyle}\\ \midrule
% 0.7514 (0.0608) & 0.7718 (0.0424) & 0.7550 (0.0553) & 0.7293 (0.0809) & \textbf{0.7721} (0.0428) 

w/o Style noise & w/o Style mixing & w/o AdvOpt & w/o AdvOpt \& Style mixing & w/o AdvOpt \& Style noise (MixStyle-DA) & \textbf{MaxStyle}\\ \midrule
0.8058	(0.0467) & 0.8295	(0.0395) & 0.8188 (0.0478) & 0.8121	(0.0516) & 0.7981 (0.0652) & \textbf{0.8321(0.0339)}
\\ \bottomrule
\end{tabular}%
}
\end{table}
\noindent\textbf{Ablation study:}
We highlight the importance of increased style diversity via composing style mixing and style noise as well as improved style hardness with adversarial training in the ablation study. Results are shown in Table~\ref{tab:abl_adversarial_training}. It can be observed that removing each component leads to performance degradation.

\section{Discussion and conclusion}
%% future work, including in compositional framework: Augmix.
We introduced MaxStyle, which is a powerful plug-in data augmentation module for single domain generalization. It attaches an auxiliary style-augmented image decoder to a segmentation network for image reconstruction and style augmentation. MaxStyle is capable of generating stylized images with improved diversity and hardness, forcing the network to learn robust features for domain generalization. From the causal perspective~\cite{Castro_2020_Nature}, the image decoder with MaxStyle can be viewed as a latent data generation process with multi-scale style interventions, conditioned on the extracted high-level causal variables (e.g., content and style). By explicitly feeding the network with images of diverse `hard' styles, the spurious correlation between image style and labels are weakened. The network thus gains the ability to resist unseen distributional shifts. We also highlighted the importance of combining both reconstructive features and discriminant features for robust segmentation (Sec.~\ref{sec:auxiliary image decoder}), which is in line with classical robust subspace methods~\cite{Fidler_2006_TPAMI_Combining}. We validated MaxStyle's efficacy on both cardiac datasets and prostate datasets across various unseen test sets, demonstrating its superiority over competitive methods under both low- and high-data training regimes. It is worth mentioning here that our method does not necessarily sacrifice the IID performance. On the prostate segmentation task, our method can significantly improve the IID performance compared to the baseline method (0.8597 vs 0.8277, see Table S3). For the cardiac segmentation task, the IID performance slightly drops (e.g, 0.8104 vs 0.8108). We hypothesize that the IID performance degradation is task and dataset-dependent. Like other adversarial data augmentation methods~\cite{Madry_2017_PGDattack, Miyato_2018_PAMI_VAT}, the main limitation
of MaxStyle is that the improved robustness comes at the cost of longer training time due to the adversarial optimization procedure, e.g. increased by a factor of $\sim1.8$ in our case. We hope that MaxStyle will enable more data-efficient, robust and reliable deep models. Integrating MaxStyle into composite data augmentation frameworks~\cite{Hendrycks_ICLR_2020_AugMix} would be an interesting direction.
\subsubsection{Acknowledgment:} This work was supported by two EPSRC Programme Grants (EP/P001009/1,
EP/W01842X/1) and the UKRI Innovate UK Grant (No.104691).
\newpage
\clearpage
\printbibliography
% % \bibliographystyle{splncs04}
% % \bibliography{mybib}
\appendix
% MICCAI-2022: Authors should not submit text materials beyond figure and table captions, definition of variables in equations, or detailed proof of a theorem. Captions should not exceed 100 words. Additionally, authors may submit supplementary videos without any identification markers. All supplementary material must be self-contained and zipped into a single file. Only the following formats are allowed: avi, doc, docx, mp4, pdf, wmv. We encourage authors to submit videos using an MP4 codec such as DivX contained in an AVI. Also, please submit a README text file with each video specifying the exact codec used and a URL where the codec can be downloaded.

\beginsupplement
% \section*{Supplementary material}
\begin{table}[!ht]
\caption{\textbf{Cardiac segmentation datasets.} \small{bSSFP: balanced steady-state free precession; LGE:late gadolinium-enhanced.}}
\label{tab:cardaic multi-face robustness test sets}
\resizebox{\textwidth}{!}{%
\begin{threeparttable}
\begin{tabular}{@{}c|c|ccccccccc@{}}
\toprule
\multicolumn{1}{l|}{} & \textbf{Single source domain} & \multicolumn{9}{c}{\textbf{Multiple unseen test domains}} \\ 
\multicolumn{1}{l|}{} & ACDC & \multicolumn{4}{c|}{ACDC w/ corruptions (ACDC-C)} & \multicolumn{4}{c|}{Cross-site datasets} & Cross-sequence \\  
Dataset & Train/val/intra-domain test & Bias field & Ghosting & Spiking & \multicolumn{1}{c|}{Motion} & Site A & Site B & Site C & \multicolumn{1}{c|}{Site D} & LGE \\ \midrule
\# subjects & \begin{tabular}[c]{@{}c@{}}Low data regime: 10/10/20\\ High data regime: 70/10/20\end{tabular} & 60 & 60 & 60 & \multicolumn{1}{c|}{60} & 75 & 50 & 25 & \multicolumn{1}{c|}{45} & 45 \\
Sequence & bSSFP & \multicolumn{4}{c|}{bSSFP} & \multicolumn{4}{c|}{bSSFP} & LGE \\
Data source & {ACDC}~\cite{Bernard_2018_TMI} & \multicolumn{4}{c|}{ACDC~\cite{Bernard_2018_TMI}+TorchIO~\cite{perez-garcia_torchio_2021}} & M\&Ms~\cite{Campello_2021_TMI_Multi} & M\&Ms~\cite{Campello_2021_TMI_Multi} & M\&Ms~\cite{Campello_2021_TMI_Multi} & \multicolumn{1}{c|}{MSCMRSeg~\cite{Zhuang_Arxiv_2020}} & MSCMRSeg~\cite{Zhuang_Arxiv_2020} \\ \bottomrule
\end{tabular}%
\end{threeparttable}
}
\end{table}

\begin{figure}[t]
    \centering
    \includegraphics[width=0.8\textwidth]{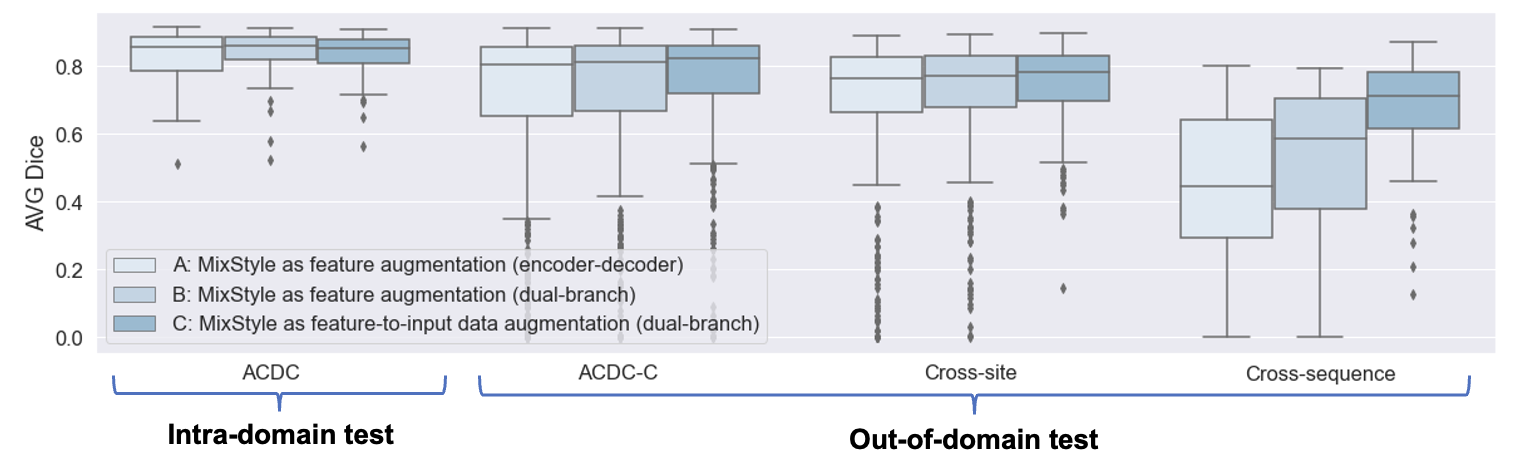}
    \caption{\textbf{Cardiac intra-domain and out-of-domain segmentation performance of networks trained with the three  approaches presented in Fig.1}. The feature-to-input space data augmentation-based (C) clearly outperforms feature augmentation based ones (A,B) on the out-of-domain test data, especially on the most challenging \emph{cross-sequence} test set. Models were trained under the cardiac low training data regime. }
    \label{fig:three_approach}
\end{figure}
\begin{figure}[!h]
    \centering
    \includegraphics[width=0.6\textwidth]{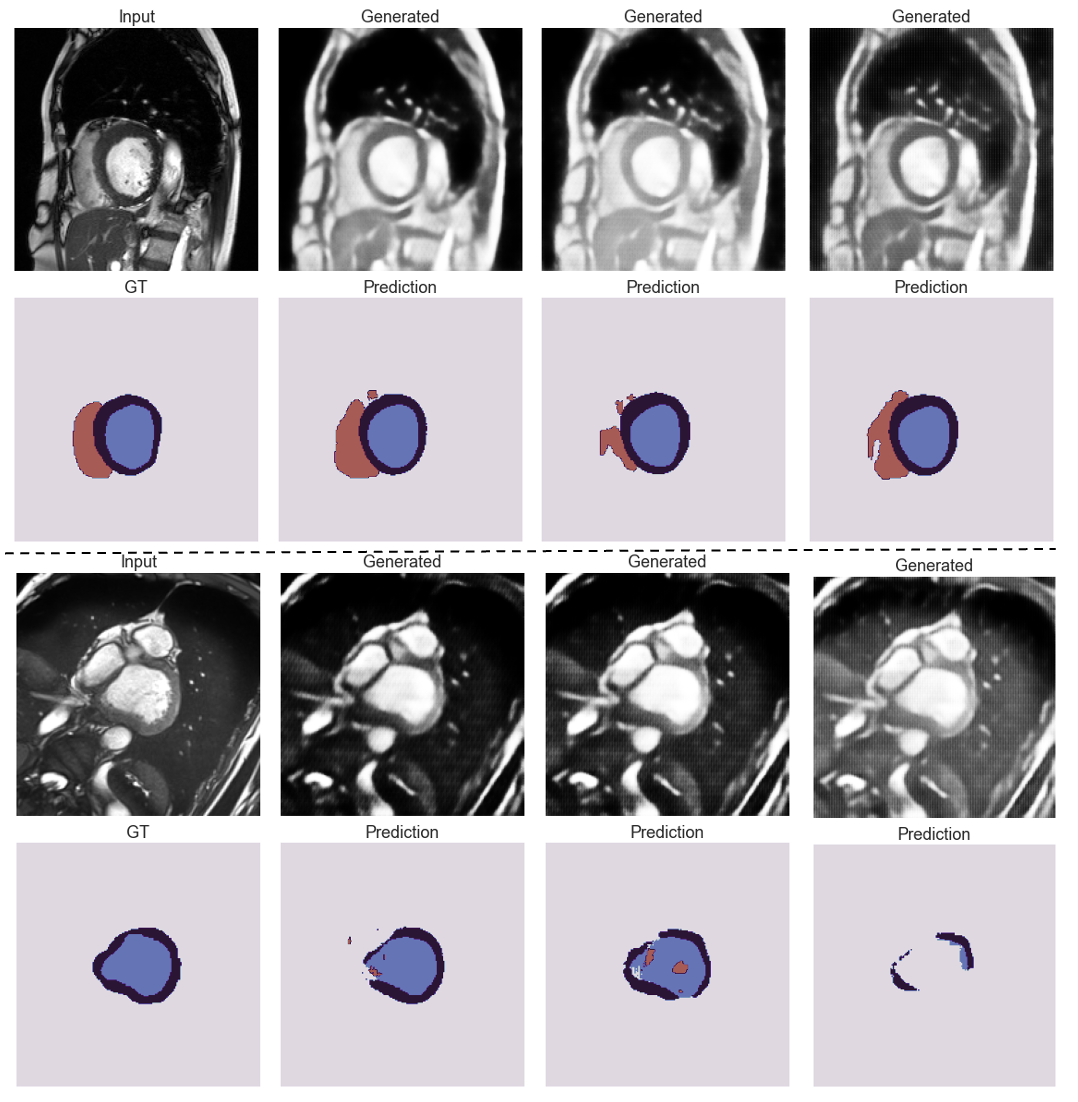}
    \caption{\textbf{MaxStyle} generates images with diverse appearances (brightness, textures) and different image quality, e.g., w/o or w/ blurring, fabric patterns to fool the segmentor to produce incorrect predictions.}
    \label{fig:generated_images}
\end{figure}
\begin{figure}[t]
    \centering
    \includegraphics[width=0.48\textwidth]{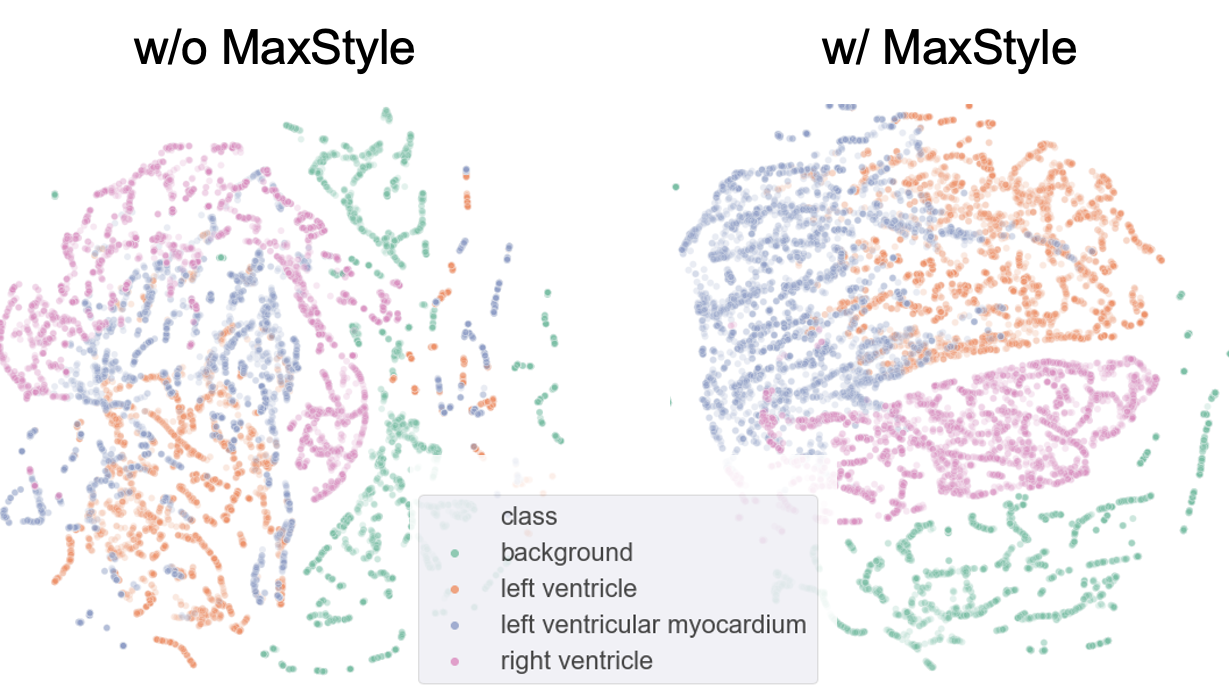}
    \caption{\textbf{t-SNE feature visualization on the \emph{most challenging} unseen cardiac test set (cross-sequence).} MaxStyle leads to well separated compact latent embeddings for different classes (in different colors), suggesting its improved robustness against unseen domain shifts. Features analyzed are from the last hidden layers of models trained under the cardiac low training data setting.}
    \label{fig:tsne}
\end{figure}
\begin{table}[t]
\centering
\caption{\textbf{Prostate segmentation datasets.} The \emph{public} prostate segmentation dataset from the \datahref{http://medicaldecathlon.com/}{Medical Decathlon challenge}~\cite{Antonelli_2021_Medical_Decathlon} is used for training and intra-domain test, where images are  all from a \emph{single} site. A \datahref{https://liuquande.github.io/SAML/}{prostate multi-site dataset}~\cite{liu2020ms,liu2020saml} collated from three independent public challenge datasets is used for cross-site robustness test. All images have been preprocessed to have the same spatial resolution: $0.625 \times 0.625 \times 3.6~\textit{mm}^{3}$, the same image size (via padding/cropping): $224 \times 224$, and the same intensity range: $[0,1]$.}
\label{tab:prostate multi-site robustness test sets}
\resizebox{0.8\textwidth}{!}{%
\begin{tabular}{@{}c|c|cccccc@{}}
\toprule
\multicolumn{1}{l|}{} & \multicolumn{1}{l|}{\textbf{Single source domain}} & \multicolumn{6}{c}{\textbf{Multi-site unseen test domains}} \\ 
Dataset & Train/val/intra-domain test & Site A & Site B & Site C & Site D & Site E & Site F \\\midrule
\# subjects & 22/3/7 & 30 & 30 & 19 & 13 & 12 & 12 \\
% Endorectal Coil & - & Surface & Endorectal & - & - & Endorectal & Endorectal \\ 
Data source & Medical Decathlon~\cite{Antonelli_2021_Medical_Decathlon} & {NCI-ISBI-2013~\cite{NCI-ISBI_2013}}  & {NCI-ISBI-2013~\cite{NCI-ISBI_2013}} & I2CVB~\cite{Lemaitre_2015_I2CVB}& \multicolumn{3}{c}{ PROMISE12~\cite{Litjens_2014_Promise_12}} \\ \bottomrule
\end{tabular}%
}\end{table}
\begin{table}[t]
\centering
\caption{\textbf{Evaluation results on the prostate intra-domain and unseen cross-site test sets.} Models were all trained using the same setting for fair comparison, e.g., the same backbone dual-branch network, the same multi-task loss with an Adam optimizer (batch size$=20$ learning rate$=1e^{-4}$, 600 epochs). }
\label{tab:prostate}
\resizebox{0.6\textwidth}{!}{%
\begin{tabular}{@{}c|c|c|cccccc@{}}
\toprule
Method & IID & OOD & \textit{A} & \textit{B} & \textit{C} & \textit{D} & \textit{E} & \textit{F} \\ \midrule
Baseline & 0.8277 & 0.7017~\textcolor{gray}{(-)} & 0.8644 & 0.5517 & 0.8240 & 0.8269 & 0.5640 & 0.5789 \\\midrule
RandConv & 0.8469 & 0.7476~\textcolor{gray}{(+7\%)} & 0.8828 & 0.6621 & 0.8554 & 0.8789 & 0.5992 & 0.6074 \\
Adv Noise & {0.8633} & 0.7325~\textcolor{gray}{(+4\%)} & 0.8939 & 0.5887 & 0.8480 & 0.8698 & 0.5877 & 0.6071 \\
Adv Bias & 0.8462 & 0.8069~\textcolor{gray}{(+15\%)} & 0.9026 & 0.7363 & 0.8195 & {0.9198} & 0.6679 & 0.7953 \\\midrule
RSC & 0.8582 & 0.7449~\textcolor{gray}{(+6\%)} & 0.8864 & 0.6224 & 0.8502 & 0.8646 & 0.6112 & 0.6343 \\
MixStyle & 0.8403 & 0.7260~\textcolor{gray}{(+3\%)} & 0.8898 & 0.5438 & 0.8154 & 0.8405 & 0.6369 & 0.6295 \\
DSU & 0.8311 & 0.7256~\textcolor{gray}{(+3\%)} & 0.8821 & 0.5924 & 0.8224 & 0.8304 & 0.5856 & 0.6403 \\\midrule
LSM & 0.8439 & 0.8209~\textcolor{gray}{(+17\%)} & 0.8941 & 0.7065 & 0.8763 & 0.9021 & \textbf{0.7827} & 0.7635 \\
% \textbf{MixStyle-DA} &  &  & & &  & && \\
% \textbf{MaxStyle } & \textbf{0.8692} & \textbf{0.8505} & \textbf{0.9069} & \textbf{0.8129} & \textbf{0.8732} & 0.9164 & \textbf{0.7506 }& \textbf{ 0.8430} \\
\textbf{MaxStyle} & 0.8597 & \textbf{0.8439}~\textbf{\textcolor{gray}{(+20\%)}} &\textbf{ 0.9054} & \textbf{0.8362} & \textbf{0.8864} & \textbf{0.9212} & 0.7160 & \textbf{0.7983} \\
\bottomrule
\end{tabular}%
}
\end{table}
\begin{table}[t]
\centering
\caption{\textbf{The effect of the number of iterations for style optimization ($n_{iter}$).} Reported values are mean (std) Dice scores over ten test (sub) sets using models trained under the high-data training setting. MaxStyle is \emph{stable} with respect to different iterations. The segmentation performance reaches the best results when performing $5$ iterations (our default setting) for style optimization.}
\label{tab:hyperparam_iterations}
\resizebox{0.7\textwidth}{!}{%
\begin{tabular}{c|c|c|c|c|c}\toprule
& $n_{iter}$=0 & $n_{iter}$=1 & $n_{iter}$=3 & \textbf{$n_{iter}$=5} & $n_{iter}$=7 \\\midrule
% DG performance (across all test domains) & 0.7550 (0.0553) & 0.7572 (0.0463) & 0.7653 (0.0423) & 0.7720 (0.0428) & 0.7642 (0.0370) \\
Mean (std) Dice scores & 0.8188 (0.0478) & 0.8256 (0.0449) & 0.8297 (0.0365) & \textbf{0.8321 (0.0339)} & 0.8317 (0.0364)\\
% Time & 6.0 & 6.6 & & 15.5 & \\
% & 5.7 & 6.2 & & 15.0 & \\
% & 6.2 & 6.9 & & 16.6 & \\
% Training time (h) & 5.9 (0.3) & 6.6 (0.3) & 8.1 (0.5) & 15.7 (0.8) & \\ 
\bottomrule
\end{tabular}}
\end{table}

% \todo{add hyper-parameters results: varying iterations, and different layers}

\end{document}